\documentclass{aastex631}

\usepackage{epsfig}

\shorttitle{HI observations of NGC 2768}
\shortauthors{Yu et al.}
\graphicspath{{./}{figures/}}

\begin{document}

\title{Deep neutral hydrogen observations of the early-type galaxy NGC 2768: collided by a newly discovered satellite galaxy?}

\author{Nai-Ping Yu}\thanks{npyu@bao.ac.cn }
\affiliation{Guizhou Radio Astronomical Observatory, Guizhou University, Guiyang 550000, China}
\affiliation{National Astronomical Observatories, Chinese Academy of Sciences, Beijing 100101, China}
\affiliation{CAS Key Laboratory of FAST, National Astronomical Observatories, Chinese Academy of Sciences, Beijing 100101, China}

\author{Ming Zhu}\thanks{mz@nao.cas.cn }
\affiliation{Guizhou Radio Astronomical Observatory, Guizhou University, Guiyang 550000, China}
\affiliation{National Astronomical Observatories, Chinese Academy of Sciences, Beijing 100101, China}
\affiliation{CAS Key Laboratory of FAST, National Astronomical Observatories, Chinese Academy of Sciences, Beijing 100101, China}

\author{Jin-Long Xu}\thanks{xujl@bao.ac.cn }
\affiliation{Guizhou Radio Astronomical Observatory, Guizhou University, Guiyang 550000, China}
\affiliation{National Astronomical Observatories, Chinese Academy of Sciences, Beijing 100101, China}
\affiliation{CAS Key Laboratory of FAST, National Astronomical Observatories, Chinese Academy of Sciences, Beijing 100101, China}

\author{Chuan-Peng Zhang}
\affiliation{Guizhou Radio Astronomical Observatory, Guizhou University, Guiyang 550000, China}
\affiliation{National Astronomical Observatories, Chinese Academy of Sciences, Beijing 100101, China}
\affiliation{CAS Key Laboratory of FAST, National Astronomical Observatories, Chinese Academy of Sciences, Beijing 100101, China}

\author{Xiao-Lan Liu}
\affiliation{Guizhou Radio Astronomical Observatory, Guizhou University, Guiyang 550000, China}
\affiliation{National Astronomical Observatories, Chinese Academy of Sciences, Beijing 100101, China}
\affiliation{CAS Key Laboratory of FAST, National Astronomical Observatories, Chinese Academy of Sciences, Beijing 100101, China}

\author{Peng Jiang}
\affiliation{Guizhou Radio Astronomical Observatory, Guizhou University, Guiyang 550000, China}
\affiliation{National Astronomical Observatories, Chinese Academy of Sciences, Beijing 100101, China}
\affiliation{CAS Key Laboratory of FAST, National Astronomical Observatories, Chinese Academy of Sciences, Beijing 100101, China}


\begin{abstract}
We present the results of a deep neutral hydrogen (H\textsc{i}) observation of the early-type galaxy NGC 2768 using the Five-hundred-meter Aperture Spherical radio Telescope (FAST). Leveraging the high sensitivity of FAST, we discover an extended gas envelope around NGC 2768. The total H\textsc{i} mass is measured to be 8.1 $\times$ 10$^8$ M$_\odot$ , representing a magnitude increase compared to previous Westerbork Synthesis Radio Telescope (WSRT) studies. Position-velocity (PV) diagram indicates the envelope mainly involves two components: an H\textsc{i} disk of NGC 2768 and a newly discovered satellite galaxy without detectable counterparts in currently deep optical surveys. The center of the gas disk is misaligned with the optical disk of NGC 2768, with more gas redshifted, indicating it has been disturbed. Our study indicates NGC 2768 is currently undergoing a transition from a spiral galaxy to an S0.  Previous deep WSRT observations reveal two dense clumps (named as ``Clumps A" and ``Clump B" throughout this paper) in the center of the envelope. We find Clump A corresponds to the densest part of the disk, while Clump B might be a newly discovered satellite galaxy which probably collided NGC 2768 about 0.38 Gyr ago. We also find tidal interactions between Clump B and PGC 2599651, NGC 2768 and UGC 4808. Based on these new findings,  we finally analyze hierarchical accretion history of NGC 2768.

\end{abstract}

\section{Introduction} 
Traditionally classified as early-type galaxies (ETGs), S0 (or lenticular) galaxies were perceived as bulge-dominated systems with quiescent stellar populations, lacking the star-forming spiral arms characteristic of late-type spirals (Hubble 1926). This view implied uniformly ancient stellar populations and minimal gas content. However, modern observations have profoundly challenged this paradigm. Integral-field spectroscopic studies, such as ATLAS\textsuperscript{3D}, reveal that up to 80$\%$ of ETGs are fast-rotating, disk-like systems with kinematic structures akin to spiral galaxies stripped of their arms (Cappellari et al. 2011; Emsellem et al. 2011). These systems often host dynamically cold, younger stellar disks embedded within older bulge components (Kuntschner et al. 2010), suggesting episodic star formation fueled by residual gas reservoirs. The coexistence of rotationally supported disks, gas reservoirs, and localized star formation in S0s undermines the classical ``quenched relic" narrative. Resolving the origins of their structural and kinematic diversity—particularly the interplay between dark matter halos, gas accretion, and feedback—remains central to understanding their role in the broader context of galaxy evolution.

NGC 2768, classified variably as an E6 elliptical galaxy (de Vaucouleurs et al. 1991) or an S0 system with a distinct outer envelope (Sandage et al. 1981; Sandage $\&$ Bedke 1994), stands as a compelling laboratory for studying galaxy evolution driven by mergers and accretion. It is the brightest galaxy in the loose Lyon Group of Galaxy 167 (Garcia 1993). Its box-like morphology and kinematic signatures—including a polar-aligned dust/gas ring (Martel et al. 2004), a rotating molecular polar disk traced by CO emission (Crocker et al. 2008), and a young stellar population (Hakobyan et al. 2008)—strongly suggest a history of accretion or minor mergers. The recent discovery of a merging dwarf satellite candidate, dubbed $``Pelops"$ (Koch et al. 2024), further supports ongoing hierarchical assembly. Notably, the detection of Type Ib supernova SN2000ds (Filippenko $\&$ Chornock 2000) and the disk-like kinematics of planetary nebulae and globular clusters (Forbes et al. 2012) hint at a possible transformation from a late-type progenitor.

Neutral hydrogen, a critical tracer of gas accretion and tidal interactions, has been detected in NGC 2768 as a tail-like structure extending ~16 kpc north of the galaxy center, with velocities slightly redshifted relative to the systemic motion (Morganti et al. 2006). While deep WSRT observations revealed faint H\textsc{i} within the optical boundaries, the limited sensitivity left open questions about the full extent, dynamics, and origin of the gas. The alignment of H\textsc{i} with the polar dust and molecular disk (Martel et al. 2004; Crocker et al. 2008) points to a shared external origin, yet the connection between the large-scale H\textsc{i}  reservoir and the galaxy’s merger history remains unclear. In this study, we present deep H\textsc{i} observations of NGC 2768 using FAST, leveraging its unparalleled sensitivity to probe low-column-density gas and resolve kinematic substructures. Our data reveal unprecedented details of the H\textsc{i}  distribution and dynamics. These results provide critical insights into the role of gas-rich mergers in shaping the multi-phase interstellar medium (ISM) of NGC 2768 and its transition from a late-type galaxy to an S0/E system.

\section{Observations and data reduction}
The observations were conducted as part of the FAST All-Sky H\textsc{i}  Survey (FASHI), an ongoing legacy survey mapping H\textsc{i}  emission across the declination range 14$^\circ$  to +66$^\circ$  (1.0-1.5 GHz) with unprecedented sensitivity. Designed to leverage FAST's unparalleled sensitivity (system temperature $<$ 20 K, effective aperture $\sim$ 300 m, pointing accuracy 7.9$^\prime$; Jiang et al. 2019, 2020), FASHI aims to establish the most complete census of extragalactic neutral hydrogen in the local Universe (z $<$ 0.09). Between August 2020 and January 2023, the survey has covered over 7,500 deg$^2$, detecting H\textsc{i} emission from more than 40,000 galaxies (Zhang et al. 2024). To probe faint circumgalactic gas around NGC 2768, we conducted six deep mapping sessions between January and December 2024 using FAST's 19-beam receiver in on-the-fly (OTF) mode. Each 2.9$^\prime$ (HPBW at 1.4 GHz) beam observed in dual polarization, achieving a system equivalent flux density of 2,000 m$^2$ K$^{-1}$. The spectral backend employed the Spec(W) spectrometer with 500 MHz bandwidth divided into 65,536 channels, yielding a velocity resolution of 1.67 km s$^{-1}$. 

Data reduction utilized the dedicated FAST pipeline HiFAST (Jing et al. 2024), which performs critical processing steps: antenna temperature calibration, baseline subtraction using iterative polynomial fitting, RFI excision via spectral kurtosis, standing-wave removal through Fourier filtering, and flux calibration tied to the standard source 3C 286. The final data cube is gridded to 1$^\prime$ pixels. Full technical details of the pipeline are documented in the HiFAST online cookbook\footnote{https://hifast.readthedocs.io/}.

\section{Results and analysis}
If we smooth the beam to spacial resolution of 9.3$^{\prime}$, an integrated H\textsc{i} flux of 1.6Jy km s$^{-1}$ is detected in the central region of NGC 2768. This value is consistent with the previous observations by the Effelsberg 100 m telescope whose half-power beamwidth is $\sim$ 9.3$^{\prime}$ (Huchtmeier et al. 1995). At a velocity resolution of 6.7 km s$^{-1}$, the noise of our data is 0.4 mJy beam$^{-1}$. At this noise level, the 5$\sigma$ column density threshold within one velocity resolution element is 6.0 $\times$ 10$^{16}$ atoms cm$^{-2}$. This value is two orders of magnitude lower than that of WSRT, enabling us to detect new fine structures in this system. In the following subsections, we will present the new findings.

\begin{figure}
\center
\psfig{file=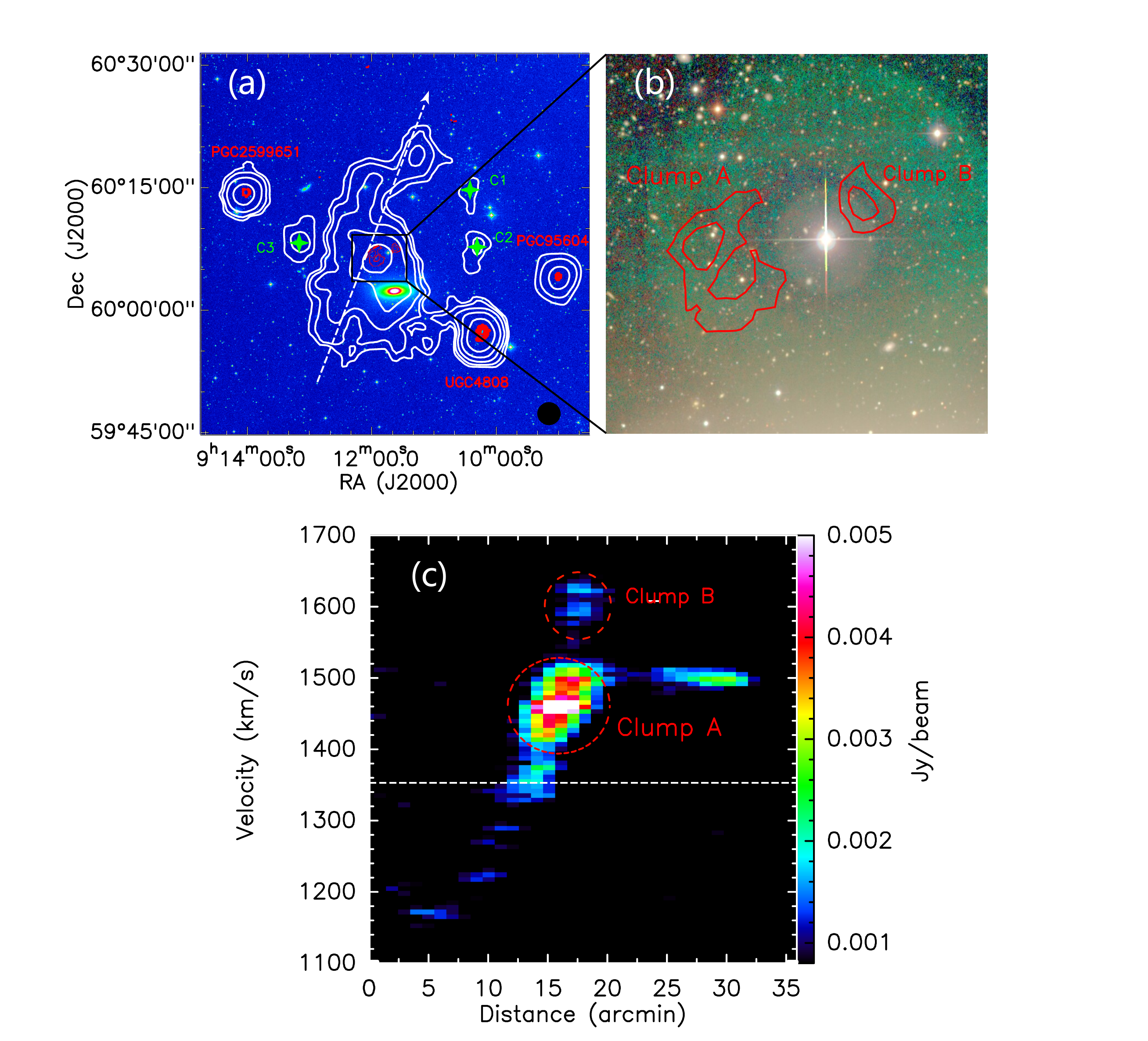,width=6in,height=5.5in}
\caption{a: The Digitized Sky Survey (DSS) $R$-band optical image of the NGC 2768, which is the most luminous galaxy in the image. The white contours are H\textsc{i} column density obtained by FAST, and levels are 1.1 $\times$ 10$^{18}$, 2.2 $\times$ 10$^{18}$, 4.4 $\times$ 10$^{18}$, 8.8 $\times$ 10$^{18}$ and 1.8 $\times$ 10$^{19}$ cm$^{-2}$. The dashed white arrow show the position and direction of the PV diagram shown in the bottom panel. The black circle indicates the FAST beam size of 2.9$^\prime$. b: H\textsc{i} surface density distribution obtained by WSRT in contours of 1.6 $\times$ 10$^{19}$  and 3.2 $\times$ 10$^{19}$ cm$^{-2}$ superimposed on the deep optical image of MATLAS. c: PV diagram of the FAST H\textsc{i} data cube cut along the direction and position shown in panel a. The white dashed line indicates the system velocity of NGC 2768.}
\end{figure}
\begin{figure}
\center
\psfig{file=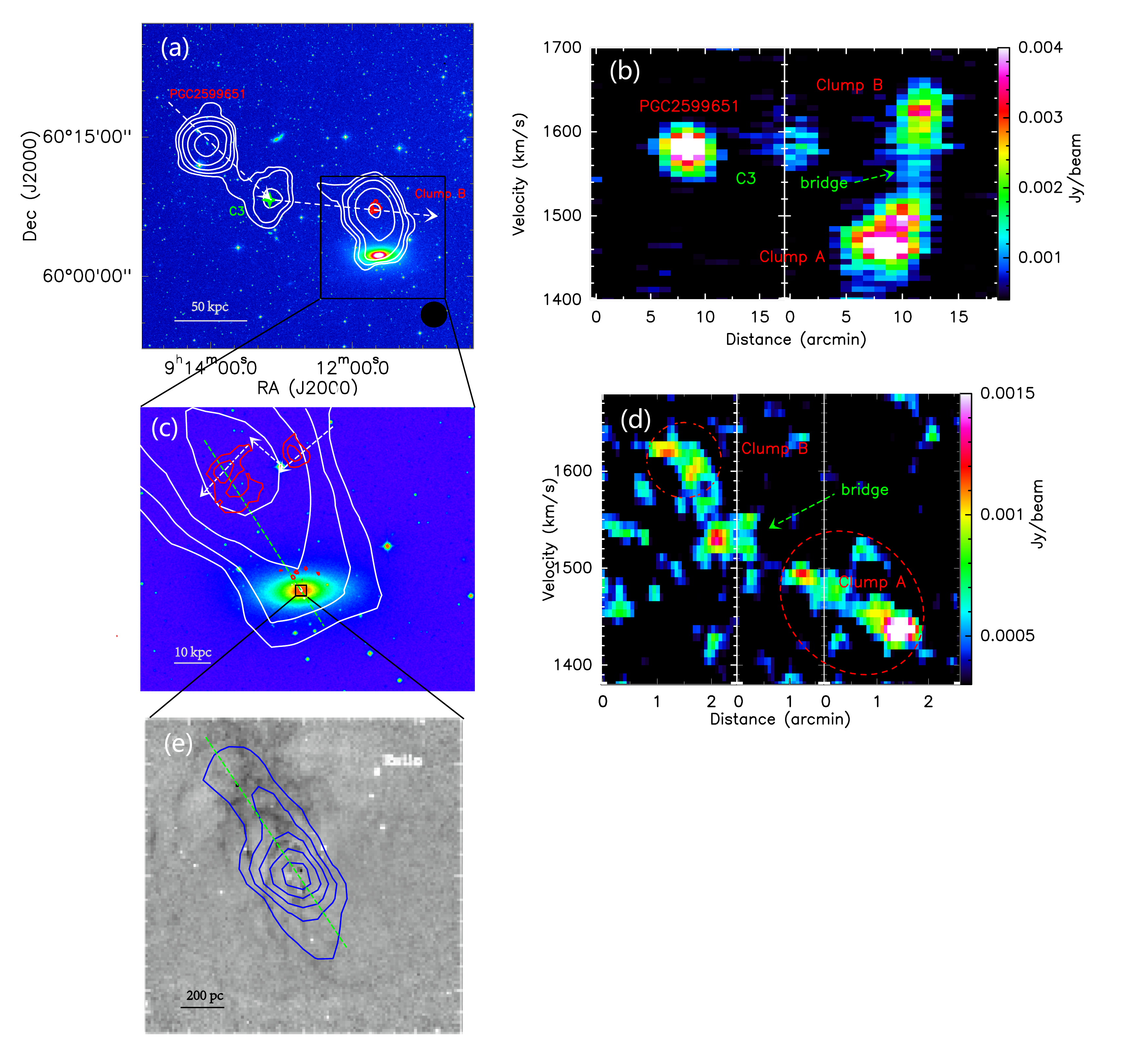,width=5.7in,height=5.7in}
\caption{The cool ISM in NGC 2768 in different scales. a: H\textsc{i} column densities of Clump B integrated over 1570-1660 km s$^{-1}$. The white contour levels are 5.5 $\times$ 10$^{17}$, 1.1 $\times$ 10$^{18}$, 2.2 $\times$ 10$^{18}$, 4.4 $\times$ 10$^{18}$ and 8.8 $\times$ 10$^{18}$ cm$^{-2}$ for FAST data, and the red contour levels are 3.0 $\times$ 10$^{19}$, 6.0 $\times$ 10$^{19}$ and 1.2 $\times$ 10$^{20}$ cm$^{-2}$ for WSRT data. The black circle indicates the FAST beam size of 2.9$^\prime$. The white arrows show the positions and directions of the PV diagram shown in the right panel. b: PV diagram of the FAST H\textsc{i} data cube cut along the directions and positions shown in the left panel. c: H\textsc{i} column density of Clump A obtained by FAST in white contours of  0.9 $\times$ 10$^{18}$, 1.8 $\times$ 10$^{18}$, 3.6 $\times$ 10$^{18}$, 7.2 $\times$ 10$^{18}$ and 1.4 $\times$ 10$^{19}$ cm$^{-2}$ superimposed on the DSS image. The red contours are Clump A and Clump B H\textsc{i} intensity observed by WSRT. Contour levels are the same as Fig.1. The dashed white arrows show the positions and directions of the PV diagram shown in the right panel. The dashed green line indicates the direction of Clump A and the CO disk of NGC 2768. d: PV diagram of the WSRT H\textsc{i} data cube cut along the directions and positions shown in the left panel. e: CO (1-0) intensity distribution in blue contours of 0.25, 0.75, 1.0 and 1.25 Jy km s$^{-1}$ superimposed on the HST dust map from Martel et al. (2004). The green dashed line is the same as panel c. }
\end{figure}

\subsection{Disturbed H\textsc{i} disk of NGC 2768}
Fig. 1 compares the 21 cm H\textsc{i} emissions of NGC 2768 from FAST and previous WSRT observations (Serra et al. 2012). While WSRT detected two dense clumps $\sim$32 kpc north of the galactic center (right panel), our deep FAST data reveals a large diffuse H\textsc{i} envelope encompassing the galaxy. The total H\textsc{i} mass is derived as:
\begin{equation}
M = 2.35 m D^2 \int N(x,y) dxdy
\end{equation}
where the factor 2.35 is the mean atomic weight, $m$ is the mass of atomic hydrogen, $D$ is the distance of 21.8 Mpc, $dx$ and $dy$ are the pixel sizes, $N(x,y)$ is the H\textsc{i} column density which could be calculated through
\begin{equation}
N (x,y) = 1.1 \times 10^{24} \frac{\int S_v dv}{\theta_{beam}^2}
\end{equation}
where $S_v$ is the flux density in unit of Jy, and $\theta$ is the beam size of FAST in arcsec. The mass of the envelope is derived to be 8.1 $\times$ 10$^8$ M$_\odot$, exceeding the WSRT-derived value of $6.5 \times 10^7 \, M_\odot$ (Serra et al. 2012) by an order of magnitude. This implies $\gtrsim90\%$ of the gas in this region resides in diffuse components ($N_{\rm H\textsc{i}} < 10^{19}\,{\rm cm}^{-2}$).

The envelope's density peak is offset by $\sim 5^\prime$ (32 kpc) from the optical disk, aligning with the WSRT clumps. The PV diagram cut through the envelope (Fig. 1) reveals two kinematic components: 
\begin{itemize}
\item An S-shaped structure spanning $1150$–$1520$ km s$^{-1}$ (centroid $1335$ km s$^{-1}$), consistent with NGC 2768's systemic velocity ($1353$ km s$^{-1}$). The rotational signature and velocity alignment identify this as the galaxy's H\textsc{i} disk.
\item A high-velocity plume ($1570$–$1660$ km s$^{-1}$) corresponding to Clump B (see Section 3.2).
\end{itemize}
From the PV diagram in Fig. 1, it can be noticed that the disk is asymmetry, with more gas redshifted, likely induced by recent interaction with Clump B. This kinematic disturbance, coupled with the disk-envelope misalignment, strongly suggests ongoing gas stripping—a key mechanism in the morphological transformation of NGC 2768 from a spiral progenitor to an S0 system. The debate of whether NGC 2768 is an elliptical galaxy or an S0 type galaxy has been \textcolor{red}{going on} for a long time. Based on the SAURON results, McDermid et al. (2006) found this galaxy to be a fast rotator without evidence for a disk, and adopt the RC3 classification of NGC 2768 as an E6. Proctor et al. (2009) used a new technique for measuring the two-dimensional velocity moments of the stellar populations of NGC 2768. Their results are consistent with the studies that NGC 2768 is an S0. Forbes et al. (2012) directly compared a sample of planetary nebulae, globular clusters and galaxy starlight velocities of NGC 2768. They find NGC 2768 resembles a spiral galaxy with arms removed. Our detection of a large H\textsc{i} envelope, the misalignment of the center of the gas envelope and optical disk of NGC 2768, and the asymmetry S-like structure in the PV diagram with more gas redshifted, support the idea that NGC 2768 has been disturbed and is currently undergoing a transition from a spiral galaxy to an S0.

\subsection{Clump B: A Dark Matter-Dominated Interacting Satellite?}
Based on the reduced WSRT standard date cube from Serra et al. (2012), we find two dense clumps in the densest part of the H\textsc{i} gas envelope (Fig. 1b). The east clump has a velocity range of 1400 to 1490 km s$^{-1}$, corresponding to the Clump A detected by FAST. While the west clump has a velocity range of 1590 to 1610 km s$^{-1}$, corresponding to the high-velocity plume (Clump B) shown in the PV diagram of Fig. 1. Thus Clump A and Clump B are two different structures. Although Clump B has already been detected by previous WSRT observations(e.g. Morganti et al. 2006; Serra et al. 2012), little analysis has been made.  In the followings, we will show it might be a satellite galaxy of NGC 2768 lacking detectable optical counterparts. Fig. 2a shows the integrated intensity map of Clump B. It can be noticed that the H\textsc{i} gas of Clump B, detected by FAST is more extended than that detected by WSRT. Its H\textsc{i} gas mass is estimated to be 7.8 $\times$ 10$^{7}$ M$_\odot$, about 14$\%$ of the gas disk mass of NGC 2768. The gas of Clump B in the south connects to the optical disk of NGC 2768, indicating gas accretion by NGC 2768. Besides Clump B, the H\textsc{i} gas of PGC 2599651 and a new H\textsc{i} cloud (names C3) have also been detected by FAST. C3 shows like a tail striped by Clump B from PGC 2599651, indicating Clump B might be a galaxy and is tidally interacting with PGC 2599651. 

To characterize the H\textsc{i} kinematics of Clump B, we performed three-dimensional tilted-ring modeling using the \textsc{TiRiFiC} software (Józsa et al. 2007) on the FAST data cube. This technique decomposes the gas distribution into concentric rings parameterized by geometrical properties  (kinematical center, inclination i, and position angle PA) and kinematic parameters (systemic velocity V$_{sys}$, velocity dispersion V$_{\delta}$ and rotation velocity V$_{rot}$). During the simulation, we fixed the kinematic center and the system velocity based on the data of WSRT, while allowing other parameters to freely converge during optimization. Fig. 3 presents a comparison between the observed H\textsc{i} structure and the best-fit model. The left panel displays the PV diagram of the observed H\textsc{i} emission along the simulated major axis from the FAST data cube. The middle panel shows the PV diagram of the best-fitting model data cube in the same direction, and the right panel is the difference between the data and the best-fitting model. The residual is relatively small suggesting that our simulation is reasonable. Through this simulation, the rotational velocity and velocity dispersion for Clump B are estimated to be 58.7 and 7.5 km s$^{-1}$, respectively. The radius (R$_{HI}$) of the HI disk is chosen as 9.5 kpc estimated from Fig. 3. The dynamical mass for Clump B could be derived to be 7.9 $\times$ 10$^{9}$M$_{\odot}$ using the formula M = (V$^2$$_{rot}$ +3$\delta$$^2$$_{v}$)R$_{HI}$ /G, where V$_{rot}$ is the rotation velocity and $\delta$$_{v}$ is velocity dispersion. 
\begin{figure}
\center
\psfig{file=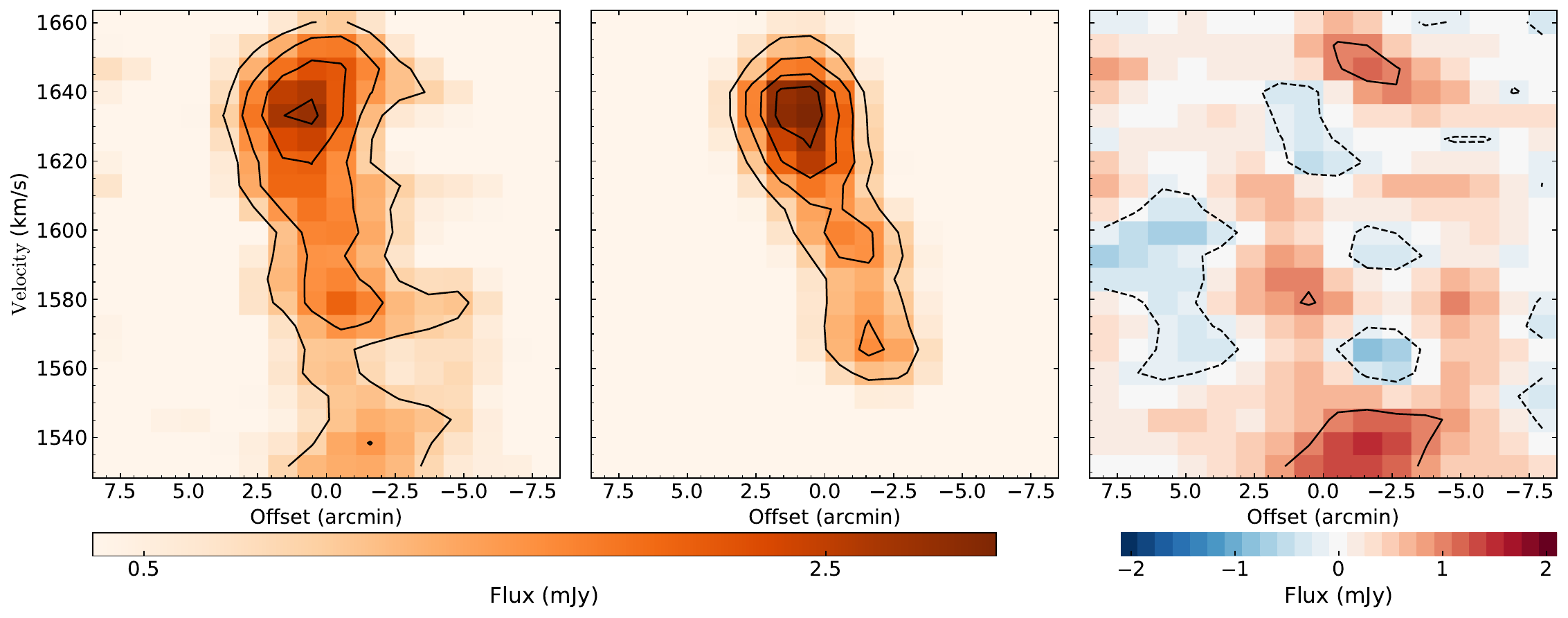,width=5.0in,height=2.0in}
\caption{Left: the PV diagram extracted from the FAST data cube in the direction of the H\textsc{i} major axies which is derived from our simulation. Middle: the PV diagram extracted from the model data cube in the same direction as the left panel. Right: the difference between observed and model data which evaluates the performance of our simulation. Contour levels are 1.0, 1.5, 2.0, and 2.5 mJy/beam for the left and middle panels. The negative and positive levels are -0.1 and 1.0 mJy/beam for the right panel.   }
\end{figure}

\begin{figure}
\center
\psfig{file=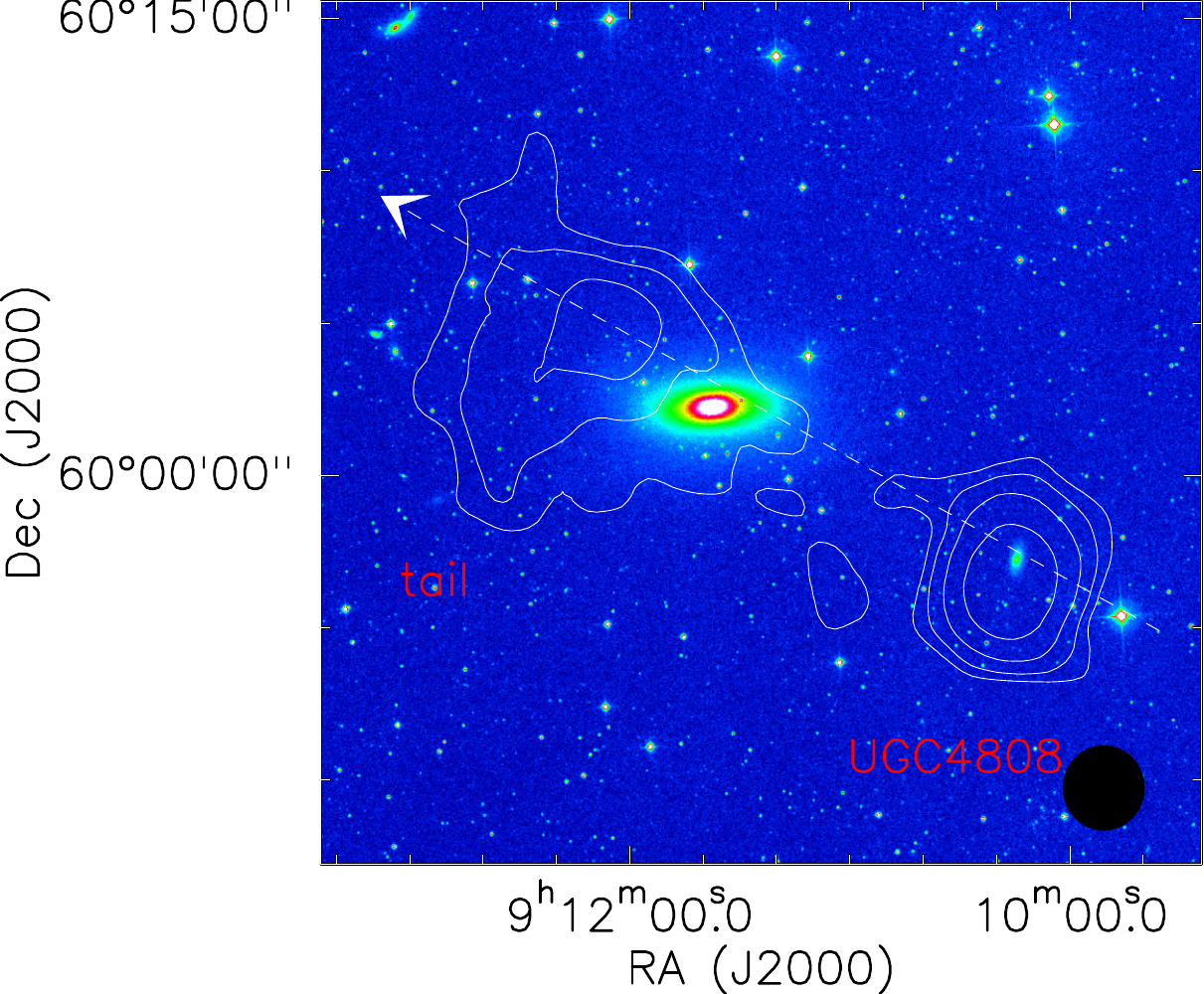,width=3.0in,height=2.5in}
\psfig{file=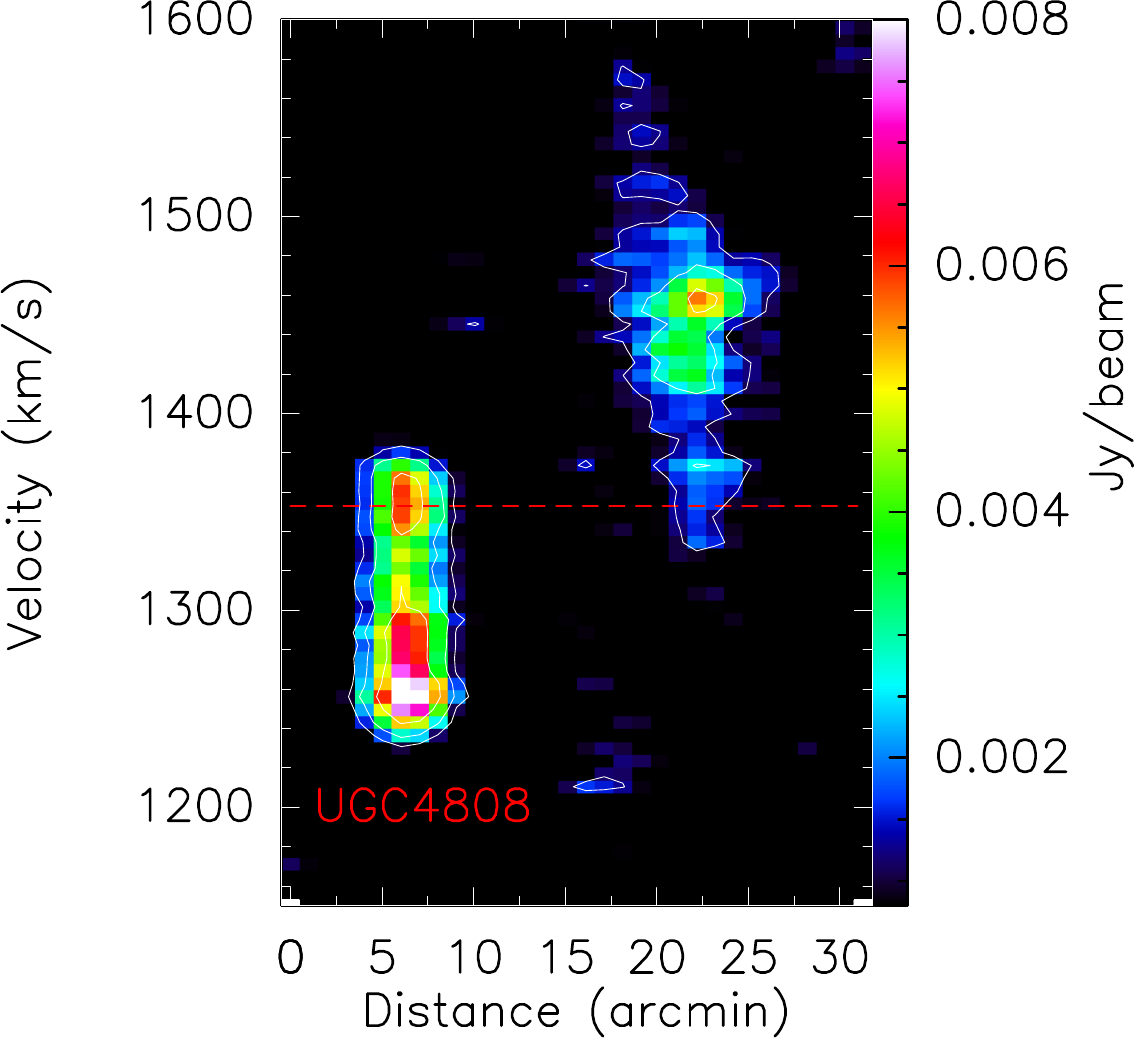,width=3.0in,height=2.5in}
\caption{Left: FAST H\textsc{i} column density of UGC 4808 integrated over 1340-1400 km s$^{-1}$. Contour levels are 6.6 $\times$ 10$^{17}$, 1.3 $\times$ 10$^{18}$, 2.6 $\times$ 10$^{18}$ and 5.3 $\times$ 10$^{18}$ cm$^{-2}$. The white arrow indicates the position and direction of the PV diagram shown in the right panel. Right: PV diagram of the FAST HI data cube cut along the direction and position shown in the left panel. Contour levels are 1.25, 2.5 and 5.0 mJy/beam. The red dashed line indicates the system velocity of NGC 2768.  }
\end{figure}

The dynamical analysis and simulations indicate Clump B might be a newly discovered galaxy in this region. We check the optical images from Sloan Digital Sky Survey (SDSS)\footnote{https://dr12.sdss.org/fields}, the Dark Energy Spectroscopic Instrument (DESI) legacy imaging surveys\footnote{https://www.legacysurvey.org/decamls/}, deep optical imaging survey of Mass Assembly of early-Type GaLAxies with their fine Structures\footnote{https://www.matlas.astro.unistra.fr/WP/} (MATLAS; Fig.1b) and ultraviolet images from the Galaxy Evolution Explore (GALEX)\footnote{https://galex.stsci.edu/GR6/}. No emissions were detected at those wavelengths. And no previously catalogued galaxy matches with this H\textsc{i} feature. A search in this region on the NED/IPAC extragalactic database (NED) also yields no findings. One possibility is that Clump B is a tidal dwarf galaxy (TDG) formed by the tidal interactions between NGC 2768 and PGC 2599651. Stars are still forming inside, making them hard to be detected. In this case, the velocity of Clump B should be the mean redshift of NGC 2768 and PGC 2599651. However, from the PV diagram of Fig.2b we notice the velocity of Clump B is both redshifted relative to NGC 2768 and PGC 2599651, suggesting that it is unlikely that  Clump B is a TDG formed by the tidal interactions between NGC 2768 and PGC 2599651. Besides, TDGs are regarded to have little dark matter. Its dark matter could be estimated by M$_{dark}$ = M$_{dyn}$ - M$_{stellar}$ - M$_{gas}$. According to the method from Zhu et al. (2021), we estimate that Clump B would not be brighter than 20.5 mag in $R$ and 21.6 mag in $B$, with the disk scale length R$_s$ = 0.74 kpc and the major-to-minor axis ratio $q$ = 0.67 as the size of Clump B is roughly 0.4$^{\prime}$ $\times$ 0.6$^{\prime}$ in the WSRT HI map. At a distance of 21.8 Mpc, we estimate that the upper limit of its luminosity is 0.26 $\times$ 10$^6$ $L_\odot$ in $R$ and 0.19 $\times$ 10$^6$ $L_\odot$ in $B$ band. Adopting the typical stellar mass-to-light ratios in the optical $R$ and $B$ bands for late-type disks or dwarf/irregular galaxies (Faber $\&$ Gallagher 1979; Portinari et al. 2004), we estimate the stellar mass (M$_{stellar}$) of Clump B would not exceed about $10^6$ $M_\odot$. Thus the dark matter of Clump B is more than 7.8 $\times$ 10$^9$ M$_{\odot}$. This mass exceeds the baryonic content by an order of magnitude, indicating Clump B is dark matter dominated. Thus, we regard that it is unlikely that Clump B is a TDG. Another possibility is that Clump B is a normal disk galaxy. Its optical stellar disk has been destroyed or merged by NGC 2768. The PV diagram of Fig. 2b displays a gas bridge between Clump A and Clump B, supporting that Clump B is interacting with NGC 2768. The higher spacial resolution of WSRT image (Fig. 1b) shows that Clump A seems to have a tail extending to Clump B. The PV diagram cut through Clump A and Clump B (Fig. 2d) also indicates Clump A is interacting with Clump B. We estimate the time since Clump B could have interacted with NGC 2768 by equation
\begin{equation}
\Delta t = \frac{\Delta R}{\sqrt{\delta^2_{group} - \Delta V^2_{sys}}}
\end{equation}
where $\Delta R$ is the projected separation (33 kpc) and $\Delta V_{sys}$ = 200 km s$^{-1}$ is the line-of-sight velocity relative to NGC 2768. The time is estimated to be 0.38 Gyr. This value is consistent with the work of Crocker et al. (2008), who estimated that NGC 2768 has been interacted by an accompany galaxy recently within 0.2 $\sim$ 0.7 Gyr. 

\subsection{Hierarchical accretions in NGC 2768}
Previous studies indicate mergers  and accretions play an important role in the history of NGC 2768 (e.g. Martel et al. 2004; Crocker et al. 2008; Zanatta et al. 2018). The optical image of Hubble Space Telescope (HST) displays a ring of dust around the central region of NGC 2768 (Martel et al. 2004). The ring lies perpendicular to the plane of NGC 2768, mostly dominant in the northern part of the galaxy. CO observations carried out by Crocker et al. (2008) also find a rotating disk perpendicularly to the major axis of NGC 2768. The CO disk is also asymmetry extending to the north-east. We show the H\textsc{i} gas of low column density detected by FAST (Fig. 2a $\&$ Fig. 2c), H\textsc{i} gas of high column density (Fig. 2c), and CO intensity distribution superimposed on the HST dust image (Fig. 2e) in different scales. A clear link between dust, CO and H\textsc{i} gas can be noticed. The cool interstellar medium (ISM) of CO and dust extend to the Clump A, Clump A is interacting with Clump B, and Clump B has a tail extending to C3 which is a tail stripped from PGC 2599651. We suggest Clump B recently collided NGC 2768. The original gas of NGC 2768 and Clump B provides the cool ISM as the accretion reservoir for the new disk perpendicular to NGC 2768. In the mean time, Clump B is also accreting material from PGC 2599651. Crocker et al. (2008) firstly noticed the link of CO, dust and H\textsc{i}. They searched all galaxies within 400 kpc  and 400 km s$^{-1}$ in relative line of sight velocity from NGC 2768, and estimated the time scale since each galaxy could have interacted with NGC 2768. They suggest the nearby galaxy UGC 4808 in the south-west is most likely to provide the cool ISM as it is the only galaxy whose interacting time with NGC 2768 is less than 0.7 Gyr. In our deep H\textsc{i} observations shown in Fig. 1a, we notice a diffuse gas bridge between NGC 2768 and UGC 4808, supporting the idea that NGC 2768 is also accreting material from UGC 4808. Fig. 4 displays the integrated intensity map of UGC 4808 and the PV diagram cut through NGC 2768 and UGC 4808. Beside the diffuse gas between them, we also notice a small tidal tail in the east of NGC 2768. In the PV diagram the contours of UGC 4808 are asymmetrical, implying tidal interactions. Our study also suggests NGC 2768 is tidally interacting with UGC 4808. However, the systemic velocity of UGC 4808 is 1295 km s$^{-1}$, blue shifted compared to that of NGC 2768. Thus UGC 4808 is unlike the main source to accelerate the gas of NGC 2768 to be redshifted. The morphologies and orientations of the dust and CO disk features also indicate gas accretions are mainly from the north-east, not from south-west. We regard in the evolutionary history of NGC 2768, it is mainly influenced by Clump B, although it is also interacted by UGC 4808. We suggest \textcolor{red}{deeper} multi-wavelength observations in this region.

\section{Conclutions}
Our deep H\textsc{i} observations of NGC 2768 with FAST have unveiled a complex gaseous ecosystem that fundamentally reshapes our understanding of this ETG’s evolutionary history. The detection of a circumgalactic H\textsc{i} envelope with a total mass of 8.1 $\times$ 10$^8$ M$\odot$—one magnitude greater than previous estimates—demonstrates FAST’s unique capability to trace low-column-density gas for studying merger remnants. Our main findings are:
\begin{itemize}
\item Morphological Transition: The gas disk of NGC 2768 has been detected for the first time. The center of the gas disk is offset from the optical disk of NGC 2768 by 32 kpc, with more gas redshifted. $\gtrsim90\%$ of the gas in this region resides in diffuse components ($N_{\rm H\textsc{i}} < 10^{19}\,{\rm cm}^{-2}$), indicating the H\textsc{i} disk has been disturbed. Our study supports the idea that NGC 2768 is currently undergoing a transition from a spiral progenitor to an S0 system.
\item A new satellite: Clump B might be a newly discovered galaxy with no detectable counterparts in currently deep optical surveys. Besides, Clump B’s high dark matter fraction rules out a tidal origin, establishing it as a primordial dwarf galaxy actively shaping NGC 2768’s evolution. The collision with Clump B ($\Delta t$ $\sim$ 0.38 Gyr) explains NGC 2768’s redshifted gas asymmetry and misaligned H\textsc{i} disk. 
\item Hierarchical accretions: The original gas of NGC 2768 and Clump B provides the cool ISM as the main accretion reservoir for the new CO/dust disk perpendicular to NGC 2768. In the mean time, Clump B is also accreting material from PGC 2599651. 
\end{itemize}
These results highlight the pivotal role of gas-rich minor mergers in regulating the morphological diversity of ETGs. Future high-resolution H\textsc{i} and stellar kinematic observations will further constrain the dark matter distribution in Clump B, while multi-wavelength campaigns targeting the NGC 2768 system could unravel the full accretion history encoded in its multi-phase ISM.

\section{Acknowledgement}
We thank the anonymous referee for the constructive comments and insightful suggestions. We thank the FAST staff for help with the FAST observations. We thank Serra P. and Crocker A. F. for sharing their WSRT and IRAM data on the web. We acknowledge supports from the National Key R$\&$D Program of China No. 2018YFE0202900. This work is supported by the Guizhou Provincial Science and Technology Projects (QKHFQ[2023]003, QKHPTRC-ZDSYS[2023]003, QKHFQ[2024]001-1, QKHJCMS[2025]015). This work is also supported by the Youth Innovation Promotion Association of Chinese Academy of Science (CAS), the National Natural Foundation of China No. 12373001, and also supported by the Open Project Program of the Key Laboratory of FAST, NAOC, Chinese Academy of Sciences. This work is also supported by the Guizhou Provincial Science and Technology Projects .

\section{Data availability}
The data underlying this article will be shared on reasonable request to the corresponding author.

\end{document}